\documentclass[aps,pra,twocolumn,showpacs]{revtex4}
\usepackage{bm}
\usepackage{psfig}

\begin{document}

\title{Improving fidelity of continuous-variable teleportation via local
operations}
\author{Jarom\'{\i}r Fiur\'{a}\v{s}ek}
\affiliation{Department of Optics, Palack\'{y} University, 17. listopadu
50, 77200 Olomouc, Czech Republic}

\begin{abstract}
We study the Braunstein-Kimble setup for teleportation of quantum state
of a single mode of optical field. We assume that the sender and receiver
share a two-mode Gaussian state and we identify optimum local Gaussian
operations that maximize the teleportation fidelity.
We consider fidelity of teleportation of pure Gaussian states and
we also introduce fidelity of the teleportation transformation.
We show on an explicit example that in some cases the optimum local operation
is not a simple unitary symplectic transformation but some more
general completely positive map.

\end{abstract}

\maketitle

\section{Introduction}

Quantum state teleportation is undoubtedly one of the most exciting
developments in the rapidly growing field of Quantum Information
Processing. In quantum teleportation, the information about the teleported
quantum state is transferred from the sender, Alice,
to the receiver, Bob, via dual classical and quantum EPR
channels \cite{Bennett93}. The latter is
established via an entangled state shared by Alice and Bob.
The teleportation protocol goes as follows:
Alice carries out a Bell-type measurement on the state she wants
to teleport and her part of the shared entangled state. She sends the result
of her measurement via classical channel to Bob, who applies  to his part of
entangled state a transformation which depends on the classical information
received from Alice.

The teleportation is perfect and Bob recovers an exact copy of the state
teleported to him by Alice only if the quantum channel is ideal
maximally entangled state. If we deal with qubits represented by polarization
states of photons, then we can employ pair of polarization-entangled photons
generated by means of spontaneous parametric down-conversion,
where the entanglement is almost perfect \cite{Zeilinger97,Boschi98}.
However, in case of continuous quantum variables
\cite{Vaidman94,Braunstein98},
an ideal EPR channel is an unphysical infinitely squeezed state.
In quantum optics, the available resource is a two-mode squeezed
vacuum state with some finite degree of squeezing
\cite{Braunstein98,Furusawa98}. Moreover,
the parts of the entangled state must be distributed among Alice and
Bob, e.g., through optical fibers. This transmission inevitably
introduces losses and noise  and the entangled state shared by Alice and
Bob will be some mixed state in general.

An important question is whether one can somehow improve the quality of
the teleportation by means of local operations on the parts of the
shared entangled state. Recently, this problem has been studied
for teleportation of qubits and it was  shown that local
transformations may indeed be helpful
\cite{Banaszek00,Badziag00,Rehacek01}. Moreover, it was demonstrated that the
optimum local transformation that maximizes the average teleportation
fidelity need not be simple unitary transformation, but some completely
positive (CP) map \cite{Badziag00,Rehacek01}. In other words,
it may be advantageous to let the parts of the shared quantum state
interact with local ancillas.

In this paper, we investigate how to improve the fidelity
of teleportation of continuous quantum variables by means of local operations
on sender's and receiver's side. The first steps in this direction were
already taken. Bowen {\em et al.} showed that in certain cases
the fidelity of teleportation of coherent or squeezed states may be improved
when Alice and Bob locally apply  squeezing transformations to their parts of
the shared quantum state \cite{Bowen01}. Kim and Lee considered
an asymmetric mixed quantum channel and showed that in that case
the fidelity of teleportation of coherent states may be enhanced
when a local transformation accompanied by decoherence is applied to
one part of the quantum channel \cite{Kim01}.

To make the problem tractable,
we restrict ourselves to the class of trace-preserving Gaussian CP
maps \cite{Demoen77,Lindblad00}. These maps preserve the Gaussian shape
of the Wigner function of the transformed state. The restriction
to Gaussian CP maps is very reasonable from the experimental point of view,
because these maps can be implemented in the laboratory
as a unitary symplectic transformation (linear canonical
transformation of quadrature operators) on the signal mode  and auxiliary
modes initially prepared in some Gaussian states.
In quantum optical setups this can be done with the help of phase shifters,
beam splitters and squeezers. Gaussian CP maps were recently applied to
description of cloning of continuous quantum variables
\cite{Lindblad00}. Another recent paper discussed the conditions under
which a given two-mode shared Gaussian state can be transformed into another
Gaussian state by means of local Gaussian CP maps \cite{Eisert01}.

The paper is organized as follows. In Sec. II we will briefly describe the
Braunstein-Kimble teleportation setup and we will derive compact
formulas for fidelities of teleportation of any pure Gaussian state.
We shall also introduce a fidelity for the teleportation operation itself.
It will turn out that this latter fidelity can be interpreted as a fidelity of
entanglement swapping. In Sec. III we will briefly review the properties of
Gaussian CP maps and we will derive optimum local Gaussian CP map
which maximizes a chosen teleportation fidelity. We shall consider two
scenarios: in the first case the transformation is applied only on one
side, in the second case both Alice and Bob may locally apply some
CP maps. In Sec. IV we present an example of our optimization procedure.
Finally, Sec. V contains conclusions.

\section{Fidelities}

We shall consider the Braunstein-Kimble  setup for
teleportation of a single mode of optical field \cite{Braunstein98}.
The quantum channel between Alice and Bob is established via two-mode
entangled state $\rho_{AB}$ fully described by its Wigner function
$W_{AB}(x_A,p_A,x_B,p_B)$. Alice mixes the mode whose state she wants to
teleport with her part of entangled state on balanced beam splitter and
she carries out a homodyne detection on each output mode thereby
measuring two commuting quadratures $X_{+}=(x_{\rm in}+x_A)\sqrt{2}$ and
$P_{-}=(p_{\rm in}-p_A)/\sqrt{2}$. After receiving the measured values of
$X_{+}$ and $P_{-}$ from Alice,
Bob displaces his part of entangled state as follows:
$x_B\rightarrow x_B+\sqrt{2}X_{+}$, $p_B\rightarrow p_B+\sqrt{2}P_{-}$.
We assume ideal homodyne detectors on Alice's side and a zero coherent
component of the entangled state $\rho_{AB}$ (mean values of all quadratures
$x_{A,B}$, $p_{A,B}$ vanish). Under these conditions the resulting state
on Bob's side possesses the same coherent component as the original state
teleported to him by Alice and the teleportation is invariant under
displacement transformation.

Ide {\em et al.} \cite{Ide01} showed that the fidelity of continuous
variable (CV) teleportation
can be improved by optimizing the gain $g$ in the modulation of the output
field whose quadratures are displaced by the amount $gX_{+}$ and $gP_{-}$.
 However, in this case the teleportation is not
in general invariant under displacement transformation and, for
instance, the fidelity of teleportation
of coherent state $|\alpha\rangle$ depends on the intensity
$|\alpha|^2$. Here we keep the gain $g$ fixed and improve the
teleportation fidelity by suitable local transformations of the shared
entangled state. Assuming fixed gain $g=\sqrt{2}$, the relation between input
and output Wigner functions of the teleported state is given by convolution
\cite{Chizhov02}
\begin{equation}
W_{\rm out}(x_2,p_2)=\int_{-\infty}^{\infty}K(x_2-x_1,p_2-p_1) W_{\rm in}(x_1,p_1)
dx_1 d p_1.
\label{WOUT}
\end{equation}
In order to express the kernel $K$ it is convenient to rewrite $W_{AB}$
as a function of the variables $x_{\pm}=x_A \pm x_B$ and
$p_{\pm}=p_A \pm p_B$,
\begin{equation}
W_{AB}(x_A,p_A,x_B,p_B)= {\cal{W}}_{AB}(x_+, p_+,x_-,p_-).
\label{WABCAL}
\end{equation}
With the help of ${\cal{W}}_{AB}$ we can write
\begin{equation}
K(x_+,p_-)= \frac{1}{4}\int_{-\infty}^{\infty}{\cal{W}}_{AB}(x_+, p_+,x_-,-p_-) dx_- d p_+ .
\label{K}
\end{equation}
In what follows we shall assume that the shared quantum state $\rho_{AB}$
is two-mode Gaussian state. This is reasonable assumption since this
class of states can be prepared in the lab.
It is computationally convenient to deal with characteristic function
of this state, defined as Fourier transform of the Wigner function,
\begin{equation}
W_{AB}({\bm r})=\frac{1}{(2\pi)^4}\int_{-\infty}^{\infty}w_{AB}({\bm q})\exp(i {\bm q}
\cdot {\bm r}) \,d^4 {\bm q} ,
\label{WABFOURIER}
\end{equation}
where ${\bm r}=(x_A, p_A, x_B, p_B)$
and ${\bm q}=(\xi_A, \eta_A,\xi_B,\eta_B)$ are real vectors.
For Gaussian state with vanishing coherent component we have
\cite{Perina91}
\begin{equation}
w_{AB}({\bm q})= \exp\left[-\frac{1}{4}  {\bm q} {\bm \Gamma}_{AB}
{\bm q}^T\right].
\label{WABGAUSS}
\end{equation}
The elements of the covariance matrix $\bm \Gamma_{AB}$  are given by
\begin{equation}
\Gamma_{AB,ij}= \langle \Delta r_i \Delta r_j\rangle +\langle
\Delta r_j  \Delta r_i \rangle,
\end{equation}
where $\Delta r_j =r_j-\langle r_j\rangle $ (note that if the coherent
component of the state vanishes then  $\langle r_j \rangle=0$).
We can express the real covariance matrix ${\bm\Gamma}_{AB}$ in terms
of three $2\times 2$ matrices $\bm A$, $\bm B$, and $\bm C$,
\begin{equation}
{\bm\Gamma}_{AB}= \left(
\begin{array}{cc}
{\bm A} & {\bm C} \\
{\bm C}^{T} & {\bm B}
\end{array}
\right).
\label{GAMMAAB}
\end{equation}
Here $\bm A$ and $\bm B$ are covariance matrices of the single modes on
Alice's and Bob's side, respectively, and $\bm C$ contains the
inter-modal correlations.

Now consider teleportation of a pure single-mode Gaussian state with
covariance matrix $\bm D$.
Since the teleportation is invariant under displacement
transformation, all states with the same covariance matrix but
different coherent components are teleported with the same fidelity. It thus
suffices to consider state with vanishing coherent amplitude,
whose characteristic function reads
\begin{equation}
w_{\rm in}({\bm q}_{\rm in})= \exp\left[-\frac{1}{4}  {\bm q}_{\rm in} \bm D
{\bm q}_{\rm in}^T\right],
\label{WINGAUSS}
\end{equation}
where ${\bm q}_{\rm in}=(\xi_{\rm in}, \eta_{\rm in})$.
Fidelity of teleportation of a pure state can be calculated as an
overlap integral of input and output Wigner functions
over the whole phase space
\begin{equation}
F= 2\pi \int_{-\infty}^{\infty}W_{\rm in}(x,p) W_{\rm out}(x,p) dx dp.
\label{FDEF}
\end{equation}
After making use of the formulas (\ref{WOUT}) and (\ref{K}),
expressing all Wigner functions as Fourier transforms of the characteristic
functions and carrying out all integrals, we arrive at a compact formula
for the teleportation fidelity,
\begin{equation}
F= \frac{2}{\sqrt{\det \bm{E}}},
\label{F}
\end{equation}
where the matrix $\bm E$ reads
\begin{equation}
{\bm E}=2 {\bm D}+ {\bm R \bm A \bm R}^T + {\bm R \bm C}+
{\bm C}^T{\bm  R}^T +{\bm B}
\label{E}
\end{equation}
and
\begin{equation}
{\bm R}=\left(
\begin{array}{cc}
1 & 0 \\
0 & -1
\end{array}
\right).
\label{R}
\end{equation}

Besides the fidelity of teleportation of certain class of Gaussian
states, one can introduce the fidelity of the teleportation process
itself. How this can be accomplished becomes clear when one notices that
the teleportation transformation (\ref{WOUT}) is a trace-preserving CP
 map \cite{Takeoka02}.
Any CP map can be  represented by positive
semidefinite operator $\chi$ on a Hilbert space which is a tensor product
of the Hilbert space of input states $\cal{H}$ and Hilbert space
of output states $\cal{K}$ \cite{Jamiolkowski72}.
This representation is not only mathematical, the state $\chi$ can be
actually prepared in the lab if we first prepare a maximally entangled
state on  Hilbert space ${\cal{H}}^{\otimes 2}$
and then apply the CP map to one part of the entangled state.
In case of CV teleportation, the maximally entangled state is the EPR
state
\begin{equation}
W_{\rm EPR}=\frac{1}{2\pi}\delta(x_1-x_2)\delta(p_1+p_2)
\label{WEPR}
\end{equation}
and the teleportation of one part of that state can
be interpreted as an entanglement swapping \cite{Loock99,Tan99}.
Hence the fidelity we obtain in this way is the fidelity of
entanglement swapping of the EPR state.

Formally, the CP map that transforms input density matrix $\rho_{\rm in}$
onto output density matrix $\rho_{\rm out}$ can be written as a partial
trace over the input Hilbert space,
\begin{equation}
\rho_{\rm out} = {\rm Tr}_{\cal H}[\chi\, \rho_{\rm in}^T \otimes
\openone_{\cal{K}}].
\label{RHOOUT}
\end{equation}
In our case, the Wigner function $W_{\rm tel}$ of the teleportation
CP map $\chi_{\rm tel}$
is closely related to the kernel $K$ because the convolution (\ref{WOUT}) is
essentially the partial trace (\ref{RHOOUT}) rewritten in terms of
Wigner functions,
\begin{equation}
W_{\rm tel}= \frac{1}{2\pi}K(x_2-x_1,p_2+p_1).
\label{WTEL}
\end{equation}
Notice the change of sign in front of  $p_1$ which reflects the transposition
in Eq. (\ref{RHOOUT}). The ideal teleportation is an identity map represented
by the EPR state (\ref{WEPR}). Now since the CP maps
are represented by positive semidefinite operators and since the ideal
transformation is represented by a pure state (\ref{WEPR}), we can calculate
the fidelity between the ideal and actual teleportation as fidelity of these
two states \cite{Raginsky01}. Thus we can write
\begin{eqnarray}
{\cal{F}}_\chi&=& 4\pi^2\int_{-\infty}^{\infty}W_{\rm EPR}(x_1,p_1,x_2,p_2) \nonumber \\
&&\qquad\quad\times  W_{\rm tel}(x_1,p_1,x_2,p_2) d x_1 d p_1 dx_2 d p_2.
\nonumber \\
\label{FSWAPDEF}
\end{eqnarray}
On inserting the explicit formulas (\ref{WEPR}) and (\ref{WTEL}) into
Eq. (\ref{FSWAPDEF}) we obtain
\begin{equation}
{\cal{F}}_\chi= K(0,0) \int_{-\infty}^{\infty}d x d p.
\label{FSWAP}
\end{equation}
We can see that $\cal{F}_\chi$ is infinite, as could have been expected since
we work in infinite dimensional Hilbert space. Nevertheless,
the fidelity (\ref{FSWAP}) can be renormalized. If we drop an infinite
constant proportional to Dirac delta function and multiply by $2\pi$,
then we obtain
\begin{equation}
{\cal{F}}=2\pi K(0,0).
\label{FSWAPREN}
\end{equation}
If we insert the explicit formula (\ref{K}) for kernel $K$
into Eq. (\ref{FSWAPREN}), then we find that
\begin{equation}
{\cal{F}}=2\pi\int_{-\infty}^{\infty}W_{AB}(x,p,-x,p) dx dp.
\label{FSWAPBLA}
\end{equation}

For Gaussian quantum channels (\ref{WABGAUSS}), this formula simplifies to
\begin{equation}
{\cal{F}}= \frac{2}{\sqrt{\det {\bm E}\bm '}}.
\label{FSWAPLAST}
\end{equation}
where the matrix $\bm E\bm '$ reads
\begin{equation}
{\bm E \bm '}=\bm R \bm A {\bm R}^T + \bm R \bm C
+ {\bm C}^T {\bm R}^T +\bm B.
\label{ETILDE}
\end{equation}
The expression for the fidelity $\cal{F}$ is
a rather special case of the formula for the fidelity of teleportation
of pure Gaussian states (\ref{F}) where we  set the covariance matrix
$\bm D$ equal to zero.
Of course, this means that $\cal{F}$ is unbounded. Nevertheless, $\cal{F}$
is a good measure of the quality of teleportation. For instance it can be
shown that ${\cal{F}}>1$ only if the state $\rho_{AB}$ is entangled
(see Appendix). In particular, if $\rho_{AB}$ is two-mode squeezed vacuum
state parametrized by squeezing constant $r$ then one gets
\begin{equation}
{\cal{F}}= \exp(2r),
\label{FSWAPSQUEEZED}
\end{equation}
hence the fidelity monotonically exponentially grows with the squeezing.

\section{Optimum local Gaussian CP map}

Our task is to maximize the fidelity of teleportation (either $F$ or
$\cal{F}$) by means of local Gaussian trace-preserving CP maps
on Alice's and Bob's side. We shall consider two scenarios:
in the first, simpler scenario the CP map is applied only on Bob's side
while in the second case both Alice and Bob may locally apply some CP maps.

Gaussian CP maps are those maps for which the Wigner function of the
corresponding operator $\chi$ has a Gaussian form.
The teleportation with Gaussian quantum channel is an example
of Gaussian CP map \cite{Takeoka02}.
The Wigner function representing single-mode trace-preserving
Gaussian CP map reads
\begin{equation}
W_\chi=\frac{1}{2 \pi^{2}\sqrt{\det \bm G}} \exp\left(
-\Delta \bm r^T \bm G^{-1} \Delta \bm r \right),
\label{WCHI}
\end{equation}
where $\bm S$ and $\bm G$ are real $2\times 2$ matrices,
moreover, $\bm G$ is symmetric positive semidefinite matrix,
\begin{equation}
\Delta \bm r = \bm r_{\rm out}- \bm S \bm r_{\rm in}^\ast ,
\end{equation}
and
\[
\bm r_{\rm out}=\left(
\begin{array}{c}
x_{\rm out} \\
p_{\rm out}
\end{array}
\right), \qquad
\bm r_{\rm in}^\ast = \left(
\begin{array}{c}
x_{\rm in} \\
-p_{\rm in}
\end{array}
\right)
\]
 are column vectors of output
and input quadratures, respectively.
Since we deal with Gaussian
states whose form is invariant under Gaussian CP maps, it suffices to
provide rule for transformation of the covariance matrix $\bm \Gamma$.
The relation beteween input and output single-mode
covariance matrices $\bm\Gamma_{\rm in}$ and $\bm \Gamma_{\rm out}$ is
given by a simple linear map \cite{Lindblad00}
\begin{equation}
\bm \Gamma_{\rm out} =\bm S\bm \Gamma_{\rm in}  \bm S^T + \bm G.
\label{CPMAP}
\end{equation}

The map (\ref{WCHI}) is completely positive
if and only if $\bm S$ and $\bm G$ satisfy an inequality
\cite{Lindblad00}
\begin{equation}
\bm G +i\bm \Sigma-i \bm S \bm \Sigma {\bm S}^T \geq 0,
\label{CPCOND}
\end{equation}
where
\begin{equation}
\bm \Sigma= \left(
\begin{array}{cc}
0  & 1 \\
-1 & 0
\end{array}
\right).
\label{SIGMA}
\end{equation}
The condition (\ref{CPCOND}) can be derived as follows:
 the Wigner function (\ref{WCHI}) must represent
a positive semidefinite operator, which imposes constraint on the
covariance matrix $\bm G$ \cite{Holevo82}. Namely, the matrix
\begin{equation}
M_{ij}=G_{ij}+[\Delta r_i,\Delta r_j],
\label{MIJ}
\end{equation}
must be positive semidefinite, where $[,]$ stands for commutator.
Making use of canonical commutation relations for the quadratures
$x_{\rm in},$ $p_{\rm in}$ and $x_{\rm out},$ $p_{\rm out}$
one arrives after some algebra at the inequality
(\ref{CPCOND}).

Assume now that a Gaussian CP map (\ref{WCHI}) is applied to Bob's part
of shared two-mode state $\rho_{AB}$. This modifies the covariance matrix
$\bm \Gamma_{AB}$,
\begin{equation}
\bm \Gamma_{AB}=\left(
\begin{array}{cc}
\bm A & \bm C\bm S^T \\
\bm S\bm C^T & \bm S\bm B\bm S^T+\bm G
\end{array}
\right),
\label{GAMMACP}
\end{equation}
The maximization of the fidelity then amounts to the
minimization of the determinant
\begin{equation}
{\cal{D}}=\det[2 \bm D+\bm R \bm A \bm R^T + \bm R\bm C\bm S^T
+ \bm S\bm C^T \bm R^T +\bm S\bm B\bm S^T+\bm G]
\label{DET}
\end{equation}
under the constraints (\ref{CPCOND}). Recall that on setting $\bm D=0$
we obtain as a special case the fidelity of entanglement
swapping (\ref{FSWAPLAST}).

We divide the optimization of the CP map into two steps. In the first
step we find optimum $\bm G$ for a given matrix $\bm S$ and then we shall
optimize over all possible matrices $\bm S$. Since the matrices $\bm S$
and $\bm G$ have altogether seven independent elements,
\begin{equation}
\bm G=\left(
\begin{array}{cc}
g_{11} & g_{12} \\
g_{12} & g_{22}
\end{array}
\right),
\qquad
\bm S=\left(
\begin{array}{cc}
s_{11} & s_{12} \\
s_{21} & s_{22}
\end{array}
\right),
\label{GSMATRIX}
\end{equation}
we have to find a global minimum of a function of seven
real variables under the constraint (\ref{CPCOND}),
which can be equivalently expressed as
\begin{equation}
g_{11}\geq 0 , \qquad g_{22}\geq 0
\label{GPOS}
\end{equation}
and
\begin{equation}
g_{11}g_{22}-g_{12}^2 -(1-s)^2 \geq 0,
\label{DETPOS}
\end{equation}
where $s=s_{11}s_{22}-s_{12}s_{21}$.
We introduce a short-hand notation for the elements of matrix
\begin{equation}
2 \bm D+\bm R \bm A \bm R^T + \bm R\bm C\bm S^T+ \bm S\bm C^T \bm R^T
+\bm S\bm B\bm S^T=
\left( \begin{array}{cc}
\alpha & \gamma \\
\gamma & \beta
\end{array}
\right).
\label{MATRIX}
\end{equation}
Notice that this matrix is, by definition, positive semidefinite,
and its elements $\alpha$, $\beta$, $\gamma$ are functions of $s_{ij}$.
Thus we can write the determinant (\ref{DET}) in a compact form,
\begin{equation}
{\cal{D}}=(\alpha+g_{11})(\beta+g_{22}) -(g_{12}+\gamma)^2.
\label{DETA}
\end{equation}

It is always optimal to choose ``extremal'' matrix $\bm G$
that satisfies the inequality (\ref{DETPOS}) as an equality. Indeed,
if a sharp inequality holds in (\ref{DETPOS}), then we can reduce the value of
diagonal elements $g_{11}$ and $g_{22}$ until the equality is reached in
(\ref{DETPOS}) and this would obviously reduce also the value of $\cal{D}$.
Hence we can write
\begin{equation}
g_{12}=\pm\sqrt{g_{11}g_{22}-(1-s)^2}
\label{GOFFDIAG}
\end{equation}
and insert into (\ref{DETA}). Furthermore, we can see that it is optimal to
choose the sign of $g_{12}$ the same as the sign of $\gamma$ and we have,
\begin{equation}
{\cal{D}}=(\alpha+g_{11})(\beta+g_{22})
-(\sqrt{g_{11}g_{22}-(1-s)^2}+|\gamma|)^2.
\label{DETB}
\end{equation}
Upon solving the set of two nonlinear extremal equations
\begin{equation}
\frac{\partial {\cal{D}}}{\partial g_{11}}=0, \qquad
\frac{\partial {\cal{D}}}{\partial g_{22}}=0,
\label{PARTIAL}
\end{equation}
we find that the optimum matrix $\bm G$ is proportional to the matrix
(\ref{MATRIX}),
\begin{eqnarray}
\bm G=\frac{|1-s|}{\sqrt{\alpha\beta-\gamma^2}}
\left(
\begin{array}{cc}
\alpha & \gamma \\
\gamma & \beta
\end{array}
\right).
\label{GOPT}
\end{eqnarray}
On inserting the elements of the optimum $\bm G$ back into Eq. (\ref{DETB})
we finally obtain
\begin{equation}
{\cal{D}}= \left(|1-s|+\sqrt{\alpha\beta-\gamma^2}\right)^2.
\label{DETC}
\end{equation}
Now $\cal{D}$ is a function of four variables $s_{11}$, $s_{22}$,
$s_{12}$ and $s_{21}$ and we have to find its  {\em global} minimum.
In general, such optimization is a hard task and can be solved only
numerically. However, we shall see that when making some assumptions
we will be able to solve this problem analytically.

It is well known that by means of local symplectic transformations it is
possible to bring any two-mode covariance matrix $\bm \Gamma_{AB}$ into
tridiagonal form \cite{Simon00},
\begin{equation}
\bm \Gamma_{AB}=
\left(
\begin{array}{cccc}
a & 0 & c_1 & 0 \\
0 & a & 0   & c_2 \\
c_1 & 0 & b & 0 \\
0 & c_2 & 0 & b
\end{array}
\right).
\label{GAMMAABBIG}
\end{equation}
It suffices to consider Gaussian quantum channels for which
all the matrices $\bm A$, $\bm B$ and $\bm C$ in (\ref{GAMMAAB}) are
diagonal. Further assume that also the covariance matrix $\bm D$ of the
teleported state is diagonal, ${\bm D}={\rm diag}(d_{11},d_{22})$.
Note that this assumption is not a serious restriction,
because any  $\bm D$ can be diagonalized by means
of reversible symplectic transformation.
In this case it can be shown that the necessary conditions on extremum
\begin{equation}
\frac{\partial {\cal{D}}}{\partial s_{12}}=0, \qquad
\frac{\partial {\cal{D}}}{\partial s_{21}}=0,
\label{PARTIALS}
\end{equation}
are satisfied when $s_{12}=s_{21}=0$. One has to be a bit
careful here because there is an absolute value in Eq. (\ref{DETC})
and three cases must be distinguished: (i) $1-s>0$,
(ii)  $1-s<0$, and (iii) $s=1$.
In all cases, the conditons on extremum (\ref{PARTIALS})
are satisfied when $s_{12}=s_{21}=0$.

We are thus lead to make the hypothesis that if the matrices
$\bm A,\bm B, \bm C, \bm D$
are diagonal, then the optimum $\bm G$ and $\bm S$ are also diagonal.
When $\bm S$ is diagonal, then $\gamma=0$, and the matrix elements $\alpha$
and $\beta$  become quadratic functions of $s_{11}$ and $s_{22}$,
respectively,
\begin{equation}
\begin{array}{c}
\alpha(s_{11})= 2d_{11}+a+2c_1 s_{11} +bs_{11}^2, \\[2mm]
\beta(s_{22})= 2d_{22}+a-2c_2 s_{22} +bs_{22}^2.
\end{array}
\label{ALPHABETA}
\end{equation}
For the sake of notational simplicity we define $x=s_{11}$ and $y=s_{22}$
and we must minimize the function
\begin{equation}
f(x,y)= |1-xy|+\sqrt{\alpha(x)\beta(y)}.
\label{FXY}
\end{equation}
The extremal equations are obtained by setting the partial derivatives
of $f(x,y)$ equal to zero,
\begin{equation}
x=\pm \sqrt{\frac{\alpha(x)}{\beta(y)}} (by-c_2),
\qquad
y=\pm \sqrt{\frac{\beta(y)}{\alpha(x)}} (bx+c_1),
\label{XY}
\end{equation}
where the signs $+$ and $-$ correspond to the cases when $1-xy>0$ and
$1-xy<0$, respectively.
From the  product of the formulas for $x$ and $y$, we can express $y$
in terms of $x$,
\begin{equation}
y=\frac{c_2(bx+c_1)}{x(b^2-1)+bc_1}.
\label{Y}
\end{equation}
Substituting this formula back into the second Eq. (\ref{XY}) and
squaring that equation, we arrive at
\begin{equation}
c_2^2 \, \alpha(x)=[x(b^2-1)+bc_1]^2\beta\left(
\frac{c_2(bx+c_1)}{x(b^2-1)+bc_1} \right).
\label{XEQUATION}
\end{equation}
This is a quadratic equation for $x$ and can be solved analytically.
In this way we identify all potential minima outside the boundary
$xy=1$. It remains to localize minima on the boundary where
$y=1/x$ and we must minimize the function
$\alpha(x)\beta(1/x)$. The condition on extremum
\begin{equation}
\frac{d}{dx} \left[\alpha(x)\beta\left(\frac{1}{x}\right)\right]=0
\label{QUARTIC}
\end{equation}
reduces to quartic equation for $x$. Upon solving this equation we get
positions of all possible minima on the boundary, i.e., we determine all
potentially optimum symplectic transformations.

Let us now consider a more general protocol, where both Alice and Bob
are allowed to apply some local Gaussian CP maps.
To make the problem tractable, we do not assume any communication
between Alice and Bob at this stage, hence they both  apply
their local operations independently.  Furthermore, we shall assume
that all relevant matrices are diagonal,
hence we shall seek the optimum two-mode CP map in the form
\begin{equation}
\bm S=\left(
\begin{array}{cccc}
u & 0 & 0 & 0 \\
0 & v & 0 & 0 \\
0 & 0 & x & 0 \\
0 & 0 & 0 & y
\end{array}
\right),
\end{equation}
\begin{equation}
\bm G=\left(
\begin{array}{cccc}
g_{A,11} & 0 & 0 & 0 \\
0 & g_{A,22} & 0 & 0 \\
0 & 0 & g_{B,11} & 0 \\
0 & 0 & 0 & g_{B,22}
\end{array}
\right).
\label{SGTWOMODE}
\end{equation}
The covariance matrix $\bm \Gamma_{AB}$ transforms according to
\begin{equation}
\bm \Gamma_{AB} \rightarrow \bm S\bm \Gamma_{AB} \bm S^T +\bm G.
\label{CPMAPTWOMODE}
\end{equation}
From  Eq. (\ref{DETPOS}) where the equality should hold and
where $g_{12}=0$, we obtain the following relations between the
elements of the optimum matrix $\bm G$,
\begin{equation}
g_{A,11}g_{A,22}=(1-uv)^2, \qquad
g_{B,11}g_{B,22}=(1-xy)^2.
\label{GABMIN}
\end{equation}
The determinant $\cal{D}$ can be expressed as
\begin{equation}
{\cal{D}}=(\alpha+g_{A,11}+g_{B,11})\left(\beta+\frac{(1-uv)^2}{g_{A,11}}
+\frac{(1-xy)^2}{g_{B,11}}\right).
\label{DETD}
\end{equation}
This function attains its global minimum when
\begin{equation}
g_{A,11}= |1-uv|\sqrt{\frac{\alpha}{\beta}}, \qquad
g_{B,11}= |1-xy|\sqrt{\frac{\alpha}{\beta}}.
\label{GABOPT}
\end{equation}
On inserting these expressions back into Eq. (\ref{DETD}), we get
\begin{equation}
{\cal{D}}=(|1-xy|+|1-uv|+\sqrt{\alpha\beta})^2,
\label{DETE}
\end{equation}
where $\alpha$ and $\beta$ are functions of four real variables
$u,v,x,y$, the elements of matrix $\bm S$. In general, the minimum of the
function (\ref{DETE}) must be found numerically. In what follows we shall
focus on the fidelity of entanglement swapping and we shall see that in
this case one can find the global minimum analytically.
The square root of the determinant (\ref{DETE}) that we must minimize
reads in this case ($d_{11}=d_{22}=0$)
\begin{eqnarray}
&&f(u,v,x,y)=|1-uv|+|1-xy| \nonumber \\
          &&\qquad +[(u^2a+2uxc_1+x^2b)(v^2a-2vyc_2+y^2b)]^{1/2}.
          \nonumber \\
\label{FUVXY}
\end{eqnarray}
This is actually a function of only three variables. This becomes
apparent when we make the following substitutions
\begin{eqnarray}
uv \rightarrow w, \qquad x\rightarrow x/v, \qquad y\rightarrow yv.
\label{SUBST}
\end{eqnarray}
The function (\ref{FUVXY}) then reads
\begin{eqnarray}
&&f(w,x,y)=|1-w|+|1-xy| \nonumber \\
        & &\quad \qquad +[(w^2a+2wxc_1+x^2b)(a-2yc_2+y^2b)]^{1/2}.
          \nonumber \\
\label{FWXY}
\end{eqnarray}
After another substitution $x=qw$ the function (\ref{FWXY})
becomes a linear function of $w$:
\begin{eqnarray}
&&f(w,q,y)=|1-w|+|1-wqy|\nonumber \\
        & &\quad \qquad +|w|[(a+2qc_1+q^2b)(a-2yc_2+y^2b)]^{1/2}.
        \nonumber \\
\label{FWQY}
\end{eqnarray}
From the linearity of (\ref{FWQY}) it is clear that the extrema are
localized at points, where one absolute value is equal to zero.
Hence we have to consider three different possibilities:

(i) $w=1$, no operation is applied on Alice's side and a CP map
is applied on Bob's side.

(ii) $wqy=1$, a symplectic transformation is applied on Bob's side.
However, this symplectic transformation can be in our case ``absorbed''
into CP map on Alice's side, hence another possibly optimum strategy is
to do nothing on Bob's side and to apply a CP map on Alice's side.

(iii) $w=0$, this means that both Alice and Bob throw away their parts of
shared quantum state and replace them with vacuum states.
Clearly, this strategy is optimum if the quantum channel is not in
entangled state, because with vacuum state at both sides
one gets maximum fidelity $\cal{F}$ obtainable without the aid of
entanglement, ${\cal{F}}_{\rm max, class}=1$.

One may object that the substitution $x=wq$ is problematic
when $w=0$ and $x\neq 0$. However, a detailed analysis reveals
that if one of the four parameters $u,v,x,y$  is set equal to zero
and the three remaining parameters are optimized,
then we once again arrive at the above listed alternatives (i)--(iii).

The strategies (i) and (ii) represent a CP map on
only one side, while nothing is performed on the other side. It was
shown above that these optimum one-sided  CP maps can be found
analytically. The search for optimum CP map would thus consist of
three parts: find one-sided optimum Gaussian CP maps on Alice's side,
on Bob's side and also consider replacement of the shared quantum state
with vacuum state and choose the optimum alternative that yields
maximum fidelity.

\section{Example of optimization}

To illustrate how the optimization works in practice,
let us assume that the covariance matrix $\bm\Gamma_{AB}$ has the
tridiagonal structure given by Eq. (\ref{GAMMAABBIG}) and the nonzero
elements read
\begin{equation}
\begin{array}{lcl}
a=1+2\sinh^2 r       & \qquad  &  c_1=-\sinh(2r), \\
b=1+2\sinh^2 r +b_0  & \qquad  &  c_2=\sinh(2r).
\end{array}
\label{abc}
\end{equation}
For $b_0=0$ we recover the covariance matrix of pure two-mode squeezed
vacuum state and  the quantum channel is in a mixed state for any $b_0>0$.

Let us analyse how the fidelity of teleportation of coherent state
can be improved by means of local transformations on Bob's side.
Since  $d_{11}=d_{22}=1$ and also $c_1=-c_2$ [c.f. Eq. (\ref{abc})],
the solution of Eq. (\ref{XEQUATION})
simplifies considerably because the functions $\alpha$  and $\beta$ are
identical. The optimum $x$ and $y$ are equal, $x=y$, and the two roots
of Eq. (\ref{XEQUATION}) read
\begin{equation}
x_1= \frac{c_2}{b-1}, \qquad  x_2 = \frac{c_2}{b+1}.
\label{XONETWO}
\end{equation}
Furthermore, the quartic equation (\ref{QUARTIC}) for optimum symplectic
transformation splits into two quadratic quations that have only two real
roots $x=\pm 1$.

\begin{figure}[!t!]
\centerline{\psfig{figure=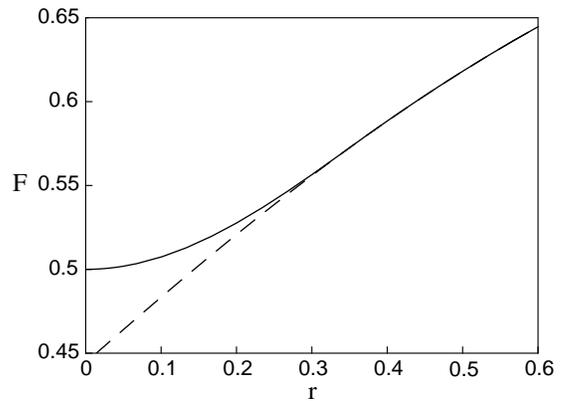,width=0.85\linewidth}}
\caption{Fidelity of teleportation of coherent state for $b_0=0.5$ and
variable squeezing $r$. The solid line shows the maximum fidelity achievable
via local CP map on Bob's side and the dashed line shows the maximum
fidelity achievable via local symplectic transformations on Bob's side.
Both curves coincide when the squeezing is higher than the
threshold $r_{\rm th}$.}
\end{figure}

We must evaluate the fidelity of teleportation of coherent state
for all these potentially optimum transformations on Bob's side,
and choose the maximum value.
The resulting fidelity is plotted in Fig. 1 for $b_0=1/2$ and a variable
degree of squeezing $r$. It turns out that if the squeezing is lower
than certain threshold $r_{\rm th}=-[\ln(1-b_0)]/2$,
then the optimum transformation on Bob's
side is a CP map with $x$ given by $x_1$ in Eq. (\ref{XONETWO}).
The parameter $x_1$ grows from zero for $r=0$ to the value $x_1=1$
that is attained when $r=r_{\rm th}$.
For higher squeezing, the best strategy is to do nothing,
i.e., the optimum operation is a symplectic transformation with $x=y=1$.
The optimum CP map for $r<r_{\rm th}$ is a simple damping process which
can be implemented with the help of a beam splitter with amplitude
transmittance $t=c_2/(b-1)$ whose two input ports are fed with Bob's part of
entangled state  and a vacuum state, respectively.
This transformation reduces the  noise represented by $b_0$
in Eq. (\ref{abc}), which in turn improves the teleportation
fidelity.

With the help of CP map on Bob's side, the fidelity of teleportation of
coherent state is always larger than the maximum fidelity $1/2$
achievable without the aid of entanglement. On the other hand,
if we allow only for unitary symplectic transformations on Bob's part of
the state, then there is a region of squeezing where the maximum
achievable fidelity is lower than $1/2$. This example clearly
illustrates that in certain cases it is advantageous to couple the
shared state to the local environment \cite{Badziag00,Rehacek01}.

\begin{figure}[!t!]
\centerline{\psfig{figure=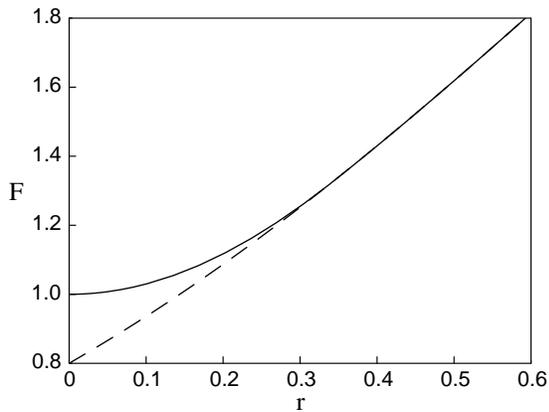,width=0.85\linewidth}}
\caption{The same as Fig. 1 but the fidelity of entanglement
swapping $\cal{F}$ is plotted.}
\end{figure}

Similar results are obtained for the fidelity of entanglement swapping
$\cal{F}$. In this case we can optimize over all
local Gaussian CP maps on both Alice's and Bob's side,
because the problem reduces to the
optimization of a Gaussian CP map on only one side, as discussed in the
previous Section. For our specific example, the optimum CP map
is actually the same as that for the fidelity of teleportation of
coherent state.
Also the dependence of the fidelity $\cal{F}$ on $r$ is qualitatively
similar to that shown in Fig. 1, see Fig. 2. In particular, there is a
region where ${\cal{F}}>1$ if Bob applies the optimum CP map,
but a restriction to local symplectic transformation results in ${\cal{F}}<1$.

\section{Conclusions}

In this paper we have shown that one can improve the fidelity of
teleportation of continuous quantum variables by means of local
operations on the sender's and receiver's parts of the shared entangled
state $\rho_{AB}$ (quantum channel). We have considered
the fidelity of teleportation of pure Gaussian states and we have also
introduced a fidelity measure for the teleportation transformation.
The latter fidelity was interpreted as a fidelity of
entanglement swapping of infinitely squeezed EPR state.
We have restricted ourselves to the class of local trace-preserving
Gaussian completely positive maps and we have shown that in this case
the optimization problem can be solved analytically.
We have demonstrated on a simple example that the optimum
local operation need not be a unitary symplectic transformation but
some more general CP map.

\vspace*{6mm}

\acknowledgments
I would like to thank R. Filip, L. Mi\v{s}ta, Jr., and J. Pe\v{r}ina
for valuable comments.
This work was supported by Grant No LN00A015 and Research Project
CEZ:J14/98 of the Czech Ministry of Education and by the EU grant under
QIPC, project IST-1999-13071 (QUICOV).

\appendix

\section{}
Here we prove that the inequality ${\cal{F}}\leq 1$ holds if the shared
quantum state $W_{AB}(x_A,p_A,x_B,p_B)$ is separable.
Our starting point is the the formula (\ref{FSWAPBLA}):
\begin{equation}
{\cal{F}}= 2\pi \int_{-\infty}^\infty W_{AB}(x,p,-x,p)d x dp.
\label{FSWAPWAB}
\end{equation}
Density matrix $\rho_{AB}$ of any separable state can be written as
a convex mixture of product states,
\begin{equation}
\rho_{AB}=\sum_j p_j \, \rho_{A,j}\otimes \rho_{B,j},
\label{SEPARABILITY}
\end{equation}
where $p_j> 0$ and
$
\sum_{j} p_j =1.
$
Formula (\ref{SEPARABILITY}) implies that
\begin{equation}
W_{AB}(x_A,p_A,x_B,p_B)=\sum_j p_j W_{A,j}(x_A,p_A)W_{B,j}(x_B,p_B).
\end{equation}
If $W_{B,j}(x_B,p_B)$ is Wigner function of the quantum state $\rho_{B,j}$,
then $W_{B,j}(-x_B,p_B)$ is a Wigner function of the quantum state
$\rho_{B,j}^T$, because the transformation $x \rightarrow -x$ and
$p\rightarrow p$ is the transposition.
For separable state (\ref{SEPARABILITY}), the formula (\ref{FSWAPWAB})
thus reduces to
\begin{equation}
{\cal{F}} = \sum_j p_j {\rm Tr}[ \rho_{A,j} \rho_{B,j}^T],
\end{equation}
where we used that
\begin{equation}
{\rm Tr}[\rho_A \rho_B]=2\pi\int_{-\infty}^\infty W_{A}(x,p)W_{B}(x,p)dx dp.
\end{equation}
The Schwarz  inequality implies that
${\rm Tr}[ \rho_{A,j} \rho_{B,j}^T] \leq 1$ and with the help of
normalization of $p_j$ we finally obtain
\begin{equation}
{\cal{F}} \leq 1.
\end{equation}
The fidelity ${\cal{F}}=1$ forms a boundary between
classical information transfer and quantum teleportation. The state
$\rho_{AB}$ must be entangled in order to achieve ${\cal{F}}>1$.
This illustrates the essential and central role of the entanglement
in the teleportation.

\end{document}